\documentclass[12pt]{iopart}
\usepackage{iopams}
\usepackage{amssymb}
\usepackage{graphicx}
\usepackage{color}
\usepackage{multicol}
\usepackage[pdftex,colorlinks,bookmarks=false,citecolor=blue,linkcolor=red,urlcolor=blue]{hyperref}

\begin{document}
\newcommand\be{\begin{equation}}
\newcommand\ee{\end{equation}}

\newcommand{\ket}[1]{|{#1}\rangle}
\newcommand{\bra}[1]{\langle{#1}|}
\newcommand{\braket}[1]{\langle{#1}\rangle}
\newcommand{\ad}{a^{\dagger}}
\newcommand{\norm}[1]{\ensuremath{| #1 |}}
\newcommand{\aver}[1]{\ensuremath{\big<#1 \big>}}
\title[Correlations and entanglement after a quench in the Bose-Hubbard model]
{Spreading of correlations and entanglement after a quench in the one-dimensional Bose-Hubbard model}
\author{Andreas M. L\"auchli}
\address{Institut Romand de Recherche Num\'erique en Physique des
  Mat\'eriaux (IRRMA), CH-1015 Lausanne, Switzerland}
\ead{laeuchli@comp-phys.org}
\author{Corinna Kollath}
\address{ 
Centre de Physique Th\'eorique, Ecole Polytechnique, CNRS, 91128
  Palaiseau Cedex, France }
\ead{kollath@cpht.polytechnique.fr}
\date{\today}
\begin{abstract}
We investigate the spreading of information in a one-dimensional Bose-Hubbard system after a sudden
parameter change. In particular, we study the time-evolution of correlations and entanglement following a quench. The investigated quantities show a light-cone like evolution, i.e. the spreading with a finite velocity. We discuss the relation of this veloctiy to other characteristic velocities of the system, like the sound velocity. The entanglement is investigated using two different measures, the von-Neuman entropy and mutual information. Whereas the von-Neumann entropy grows rapidly with time the mutual information between two small sub-systems can as well decrease after an initial increase. Additionally we show that the static von Neuman entropy characterises the location of the quantum phase transition. 
\end{abstract}

\maketitle
\section{Introduction}
Entanglement and correlations are important properties of quantum systems. The interest in these quantities is manifold. Correlations have been used since a long time to characterize the low energy properties of quantum many body states. In particular in the study of quantum phase transitions, correlations played a leading role. Entanglement has been the resource for quantum computing and quantum cryptography. Only recently entanglement has also become a major tool to understand and characterize
ground states of interacting quantum many body systems. Driven by these applications the study of entanglement has become an active field of research and many different measures have been proposed in quantum many body systems~\cite{AmicoVedral2008}. 

 Recently the interest in the fundamental questions of quantum dynamics was reinforced by the experimental realization of strongly interacting gases~\cite{BlochZwerger2007}.
In these systems the precise and rapid tunability of the system parameters and the very good decoupling from the environment open the possibility to study the quantum evolution of a system far from equilibrium. In this context the following questions arise: How does a quantum system react to a parameter change? How fast can information propagate through the system? How do correlations and entanglement between different parts of the system build up?

These and similar questions have been asked in different situations. 
Lieb and Robinson found for one-dimensional spin systems that the quantum dynamics generated by local operators obeys a light-cone like evolution, i.e.~the information spreads with a finite velocity~\cite{LiebRobinson1972}. They proved that only exponentially small corrections exist outside this light-cone in the considered systems. Many generalizations of this Lieb Robinson theorem have been developed over the last years (see Ref.~\cite{CramerEisert2008} and references therein). Further the finding of the spreading of correlations with a finite velocity after a global quench has been verified in specific integrable models~\cite{IgloiRieger2000, RigolOlshanii2006a, Cazalilla2006,CubittCirac2007, CramerOsborne2008}, as well as for critical systems which can be described by conformal field theory \cite{CalabreseCardy2006, CalabreseCardy2007}.

Motivated by the experimental investigation of a sudden parameter change across the superfluid to Mott-insulator transition \cite{GreinerBloch2002} we investigate here the time-evolution of a one-dimensional non-integrable Bose-Hubbard system after a sudden quench of the interaction
strength. For this model the recent generalization of the Lieb-Robinson theorem of Cramer et al.~\cite{CramerEisert2008} applies deep in the superfluid regime. However, no generalization exists in the presence of sizable interaction strength between the bosons. 

We start our investigation with the static properties of the von Neumann entanglement entropy in the Bose-Hubbard model. In particular we show that previous predictions from conformal field theory \cite{HolzeyWilczek1994,CalabreseCardy2004} are in very good agreement with our numerical results in the critical superfluid phase assuming a central charge $c=1$. We further point out how the knowledge on the dependence of the entanglement entropy on the system and block length in the critical phase can be used to locate the quantum phase transition.

We then turn to the discussion of the time-evolution of the quantum system after a sudden parameter change. Hereby we mainly consider a quench from the superfluid to the Mott-insulating phase. However, we also show some examples of quenches inside the superfluid or the Mott-insulating regime.  
In a previous work \cite{KollathAltman2007} we focussed on the relaxation of the system to a
quasi-stationary state after a quench. Here in contrast
we discuss the short time behaviour in detail. We study the spreading of information by the time-evolution of correlations, the von Neumann entropy, and the mutual information. 
We find that for onsite quantities the time-scale of
relaxation is set by the
hopping time of the bosons. In contrast for longer range correlations, like the
one-particle density matrix or the density-density correlations, a front spreads with almost constant
velocity which causes a `light-cone' like evolution. The
front travels at a finite speed, such that in an infinite system equilibrium
can never be reached. 
We discuss the time evolution of the von Neumann entropy of different contiguous blocks of length $l$. We find that the von Neumann entropy of a certain block shows a sharp linear rise for short time and saturates for longer times. To a good approximation the rate of the 
block entropy increase is shown to follow a boundary law, while the entropy value at saturation 
depends only on the block volume. Our results nicely corroborate recent analytical 
results \cite{CalabreseCardy2005,EisertOsborne2006,BravyiVerstraete2006}.  
In contrast to the von Neumann entanglement, the mutual information between two contiguous blocks of different length after a quench to a Mott-insulating regime drops substantially in time
compared to the values in the initial superfluid state. So for the considered quench the growth of the entropy and the loss of correlations go hand in hand.

\section{Model and methods} 
Ultracold bosons subjected to an optical lattice can be described by the Bose-Hubbard
model
 \cite{JakschZoller1998, Zwerger2003,FisherFisher1989}, here in its one-dimensional form
 \begin{equation}
\label{eq:bh}
H(J,U)= -J \sum_{j} \left(b_j^\dagger b^{\phantom{\dagger}}_{j+1}+h.c.\right)  + \frac{U}{2} \sum _{j} n_j (n_j-1),
\end{equation}
where  $b^\dagger_j$ and $b_j$ are the boson
creation and annihilation operators, and ${n}_j= b^\dagger_j
b^{\phantom{\dagger}}_j$ the number operators on site $j$.
The first term in \Eref{eq:bh} models the kinetic energy of the atoms and the parameter $J$ is
called the hopping parameter. The second term stems from the short range 
interaction between the atoms and the parameter $U$ characterizes its
strength. 
In the experimental setup of ultracold bosons in an optical lattice the parameters $U$ and $J$ 
can be directly related to the experimental parameters. In particular, by varying the lattice height
in the experiment the parameter $U/J$ can be changed by several orders of
magnitude and by anisotropic lattices the dimensionality of the system can be varied.
In equilibrium at integer filling the Bose-Hubbard model shows a quantum phase transition at a critical
value of $u_c:=(U/J)_c$ between a superfluid state ($U/J<u_c$), in which the atoms are
delocalized, to a Mott insulating state ($U/J>u_c$) in which the atoms are
localized \cite{FisherFisher1989}.  
We use $\hbar =1$ and $a=1$, where $a$ denotes the lattice spacing, throughout
the work. 
The theoretical tools we use to treat the ground state and the dynamic properties of the Bose-Hubbard model are analytical approximations and numerical methods. Static results for the entanglement entropy are obtained using the density matrix renormalization group method (DMRG) \cite{White1992, Hallberg2006, Schollwoeck2005} with up to 1000 states. The results for time-evolution are calculated using the exact diagonalization (ED) based on Krylov methods \cite{park:5870,manmana-2005-789} 
and the adaptive time-dependent DMRG (t-DMRG)
\cite{Vidal2004,WhiteFeiguin2004, DaleyVidal2004}. The ED is used to
study one-dimensional chains with periodic boundary conditions up to 16 sites,
 while the adaptive t-DMRG
is used for one dimensional systems with open boundaries and with up to 64 sites keeping several
hundred DMRG states.

\section{Static von Neumann Entanglement Entropy}
\label{sec:static}

\begin{figure} [ht]
\begin{indented}
\item\includegraphics*[width=0.8\linewidth]{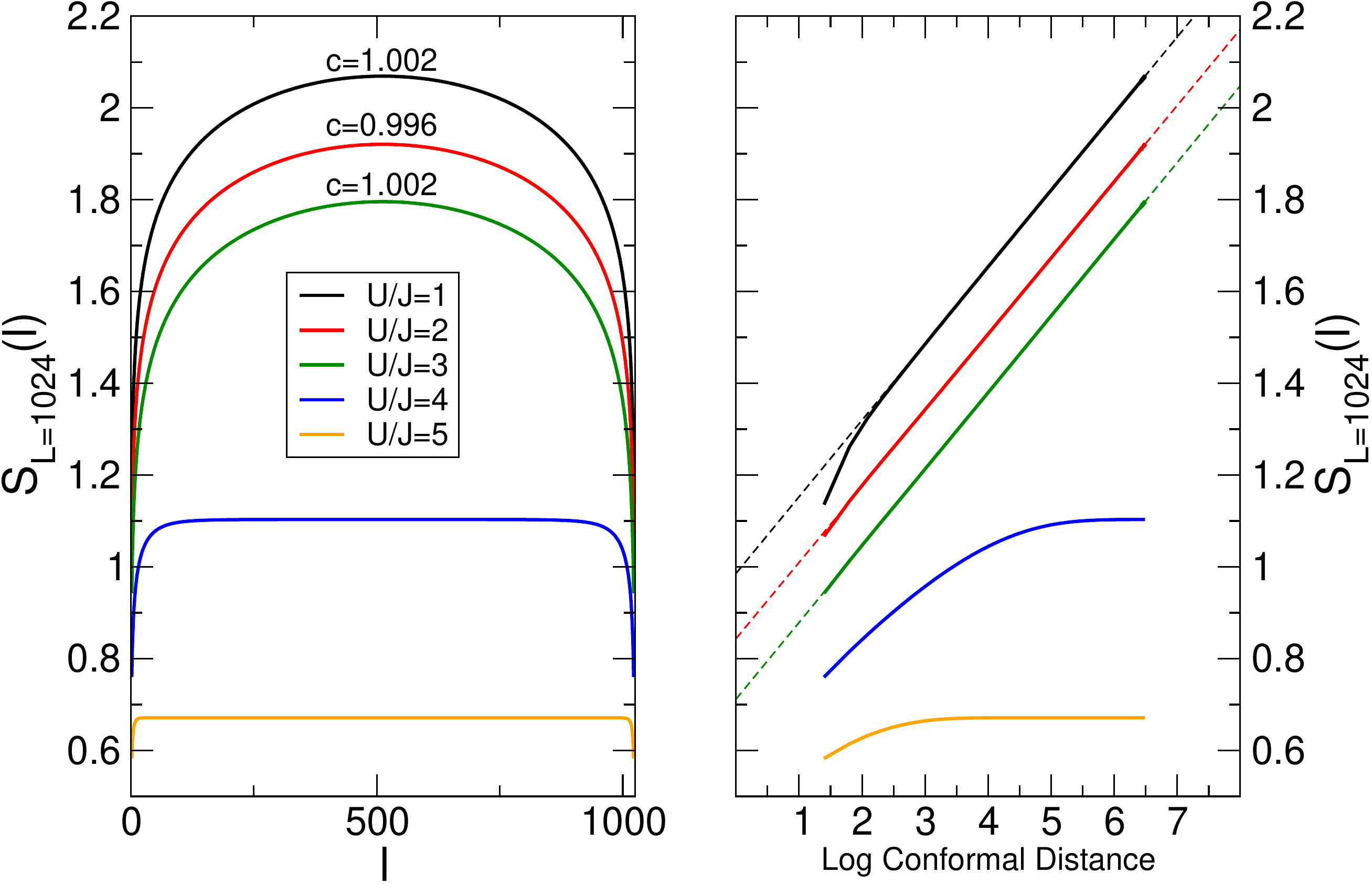}
\end{indented}
\caption{
   Left panel: The static von Neumann block entropy $S_L(l)$ for an open system of length $L=1024$ and a range of
   different interaction values $U/J$, located in the superfluid and the Mott insulating phases. The average filling is $n=1$.
   Right panel: the same data as a function of the logarithmic conformal distance $\lambda:=\log \left[(2L/\pi)\sin(\pi l/L)\right]$.
   The linear behavior of the curves of $U/J=$ 1, 2 and 3 reveals the $c=1$ critical theory. The rapid saturation of the entropy for
   $U/J=$ 4 and 5 is a consequence of the short correlation length in the Mott insulating phase.
  } 
\label{fig:entropystat}
\end{figure} 

Before we discuss the behaviour of the system after a sudden change of the parameters we would like to discuss the static properties of entanglement in the Bose-Hubbard model. In recent years many different measures have been proposed for the entanglement in a system \cite{AmicoVedral2008}.
 One of the proposed measures is the von Neumann block entropy. Let A be
a block of length $l$ and B be the remaining system. Then the von Neumann entropy
of a block A is defined by
$S_A=-\Tr_A\rho_A \log \rho_A$, where $\rho_A=\Tr_B \rho$ is the reduced density matrix of
the block A and $\rho=\ket{\Psi}\bra{\Psi}$ the pure-state density matrix of the whole
system. $\Tr_X$ denotes the trace over X.

In the superfluid phase the low
energy physics of the one-dimensional Bose-Hubbard model can be described by a Luttinger liquid, which is 
a conformal field theory with central charge $c=1$. For such a 1+1 critical system the von Neumann 
entropy has been derived  for different geometries of the subsystems A and B 
\cite{HolzeyWilczek1994,CalabreseCardy2004}. If A is a single block of length $l$ in a periodic system of 
length $L$ the von Neumann entropy is given by
\begin{equation} 
S_A=\frac{c}{3} \log \left[\frac{L}{\pi }\sin\left(\frac{\pi l}{L}\right)\right] +s_1.
\label{eq:S_pb}
\end{equation}
Here $s_1$ is a non-universal constant.
In a system with open boundary conditions which is divided at some interior
point into an interval A of length $l$ and its complement the von Neumann
entropy is 
\begin{equation} 
S_L(l)=\frac{c}{6} \log \left[\frac{2L}{\pi } \sin\left(\frac{\pi l}{L}\right)\right] + \log g +s_1/2  .
\label{eq:S_ob}
\end{equation}
 Here $\log g$ is
the boundary entropy of Affleck and Ludwig \cite{AffleckLudwig1991}.
Note that an oscillating correction term beyond the conformal field theory prediction \eref{eq:S_ob} has been found for open boundary conditions in the critical phase of the $S=1/2$ XXZ spin model \cite{LaflorencieAffleck2006}. This is a specific example of Friedel-like oscillations decaying away from the boundaries. These, in turn, lead to an oscillating, algebraically decaying correction term to the block entropy. 
In the case of the Bose Hubbard model considered here, which is not particle-hole symmetric, the open boundaries can induce a slightly non-uniform particle density, but the amplitude of the difference with respect to the nominal density decays rapidly away from the boundaries. The effect is most pronounced for small $U/J$ and becomes negligible for large $U/J$. This inhomogeneous particle density distribution is therefore expected to affect the entanglement entropy on small open systems.

For stronger interactions the quantum phase transition to the Mott-insulating phase takes places. The Mott-insulating state is characterized by an energy gap above the ground state and has a finite correlation length $\xi$. For gapped systems with a finite correlation length $\xi$, the block entropy is expected to saturate at a finite value $S_L(l) \sim \log (\xi)$ for $l\gg\xi$~\cite{VidalKitaev2003,CalabreseCardy2004}.

We show in Fig.~\ref{fig:entropystat} the von Neumann entropy of the static Bose-Hubbard model with open boundary conditions for different interaction strengths $U/J$ obtained using DMRG calculations on a system of length $L=1024$. In the left panel we show the entropy as a function of the block length $l$, while
in the right panel we rescale the $x$-axis according to the logarithmic conformal distance $\lambda:=\log \left[(2L/\pi )\sin(\pi l/L)\right]$. Eq.~\ref{eq:S_ob} then 
simplifies as $S_L(\lambda)=c/6\ \lambda + \log g + s_1/2$. 

For small interaction strength $U/J=1,2,3$ our results agree very well with the prediction \eref{eq:S_ob} of the conformal field theory for large block length $l$.
For $U/J=1$ deviations at small $l$ can be seen. They are due to the slightly inhomogeneous density in the open system. 
Linear fits to the curves in the right panel for the larger values of $\lambda$ yield slopes which are very close to the expected value of $1/6$.
The extracted value of $c$ for the three critical $U/J$ values is shown close to the corresponding curve in the left panel. The values of $c$ are very
accurately 1. 
The constant contribution $\log g + s_1/2$ becomes smaller for increasing $U/J$ due to the reduced onsite number fluctuations.
For $U/J=$ 4 and 5 a very clear saturation of the entropy as a function of the block length $l$ can be seen, revealing  the finite correlation length in the Mott insulator.
The entanglement entropy vanishes completely in the limit of $U/J \rightarrow \infty$ when the groundstate becomes a trivial product wave function of
singly occupied sites.

\begin{figure} [ht]
\begin{indented}
\item\includegraphics*[width=0.8\linewidth]{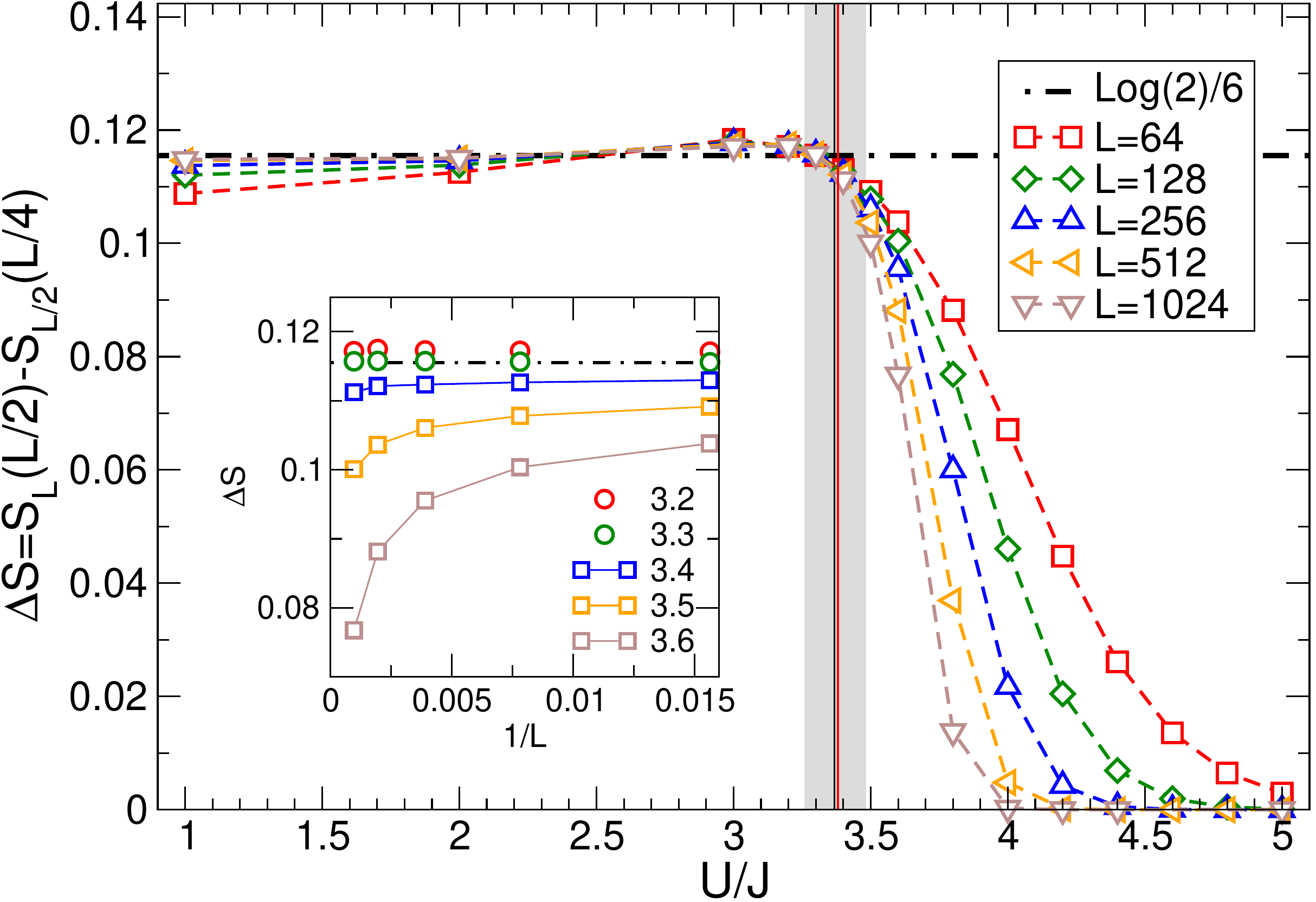}
\end{indented}
\caption{ 
   Difference $\Delta S(L)=S_L(L/2)-S_{L/2}(L/4)$ versus $U/J$ for different system lengths $L$ at average filling $n=1$. For $U/J<u_c$ 
   the expected scaling behaviour $\Delta S=c/6 \log 2$
   with $c=1$ is seen. In the $L \rightarrow \infty$ limit the data converges to a step function located at $u_c$.
   The vertical line shows the previously obtained critical values $u_c\approx 3.37(12)$~\cite{KuehnerMonien2000} (black line with grey uncertainty) and 
$u_c\approx 3.380(4)$  
    \cite{ZakrzewskiDelande2008} (red line) for comparison. 
    The deviation of small system length for $U\le 2J$ stems from inhomogeneous density distribution due to the open boundaries and is not captured in the CFT 
    approach. 
    Inset: $\Delta S$ as a function of $1/L$ for selected values of $U/J$ close to the phase transition.
}
\label{fig:entropyscal}
\end{figure} 

In the following we investigate whether it is possible to locate the critical value $u_c$ purely based on the properties of the von Neumann entropy.
Earlier high precision work with DMRG mostly used a priori knowledge about the Kosterlitz-Thouless transition to accurately locate the critical 
value~\cite{KuehnerMonien2000, ZakrzewskiDelande2008}. A recent investigation based on quantum information related quantities such as the
fidelity and single site entropies and their derivatives arrived at the conclusion that the single site entropy would not allow to locate the critical value $u_c$ 
precisely~\cite{BuonsanteVezzani2007}.

In order to locate the critical point we devise the quantity 
\begin{equation}
\Delta S(L):=S_L(L/2)-S_{L/2}(L/4)
\end{equation}
the increase of the entropy at the mid-system interface
upon doubling the system size. In a region of parameter space described by a conformal field theory with central charge $c$ one expects simply 
$\Delta S=c/6\ \log 2$ based on \Eref{eq:S_ob}. On the other hand in a gapped region with a finite correlation length one obtains $\Delta S=0$
for $L\gg \xi$, because the entropy saturates for block lengths $l$ larger than the correlation length $\xi$. As the system size increases one
therefore expects $\Delta S$ for the one-dimensional Bose-Hubbard model to scale to a step function as a function of $U/J$. 

In \Fref{fig:entropyscal} we display $\Delta S$ as a function of $U/J$ for different system sizes $L$ from 32 up to 1024 sites. In the Luttinger liquid
regime at small $U/J$ the curves for different system sizes nicely converge towards the expected value of $1/6 \log 2$. For larger $U/J$ values
the $\Delta S$ curves scale to zero for increasing $L$, indicative of the gapped phase. Based on our available data we can safely infer that the
point $U/J=3.4$ is already in the Mott insulating phase, while $U/J=3.3$ is still critical, see inset of \Fref{fig:entropyscal}. This result is in full agreement with the previously obtained 
critical values $u_c\approx 3.37(12)$~\cite{KuehnerMonien2000} and $u_c\approx 3.380(4)$~\cite{ZakrzewskiDelande2008}, based on fits of the
decaying bosonic Green's function. In future studies using a fine grid of $U/J$ values one could perform a finite size scaling of the inflection 
point to extract an even more accurate value of $u_c$. So we conclude that an appropriate scaling plot of the von Neumann entanglement entropy provides competitive 
results on the location of the quantum phase transition in the one-dimensional Bose Hubbard model without relying on a priori knowledge of the 
Kosterlitz-Thouless nature of the transition.

\section{Description of the parameter quench}

We implement the quench by an abrupt
change of the interaction strength from an initial value $U_i$  to a final
value $U_f$ at fixed hopping parameter $J$ at time $t=0$. In most cases we start from a superfluid phase ($U_i/J<u_c$) and change to the Mott-insulating regime ($U_f/J>u_c$). 
The initial wave function $\ket{\psi_0}$ for $t\le 0$ is the ground state of the
Hamiltonian $H_i=H(J,U_i)$. We investigate its time evolution for times $t>0$ subject to the
Hamiltonian $H_f=H(J,U_f)$. The time evolution of the wave function is governed  by the Schr\"odinger equation, i.e. $$\ket{\Psi(t) }= \exp\left(-\textrm{i}H_f t\right) \ket{\Psi_0}.$$ The time evolution of expectation values of relevant operators can be expressed as
\begin{equation}
\langle \hat{O}(t)\rangle = \sum_{m,m'} c^*_m c^{\phantom{*}}_{m'}
\exp\left[{-\textrm{i}(E_m-E_{m'}) t}\right] \bra{m} \hat{O} \ket{m'}.
\end{equation}
Hereby $\ket{m}$ are the eigenstates of the final Hamiltonian $H_f$ and $E_m$ the
corresponding energy eigenvalues. The expansion of the
initial state $\ket{\psi_0}$ into the eigenstates of the final
Hamiltonian, i.e.~$\ket{\psi_0}= \sum_{m}c_m \ket{m }$ with the weights $c_m$, is used.
 In a realistic case the tunneling between lattice 
sites and the interaction strength are non-zero and to determine the
time-evolution of the wave function is not an easy task. We calculated the time-evolution numerically using ED and DMRG methods and analytical approximation in certain limits.

\section{Light cone effect in correlations}
\label{sec:light_cone}
\begin{figure} [ht]
\begin{indented}
\item
\includegraphics[width=0.8\linewidth]{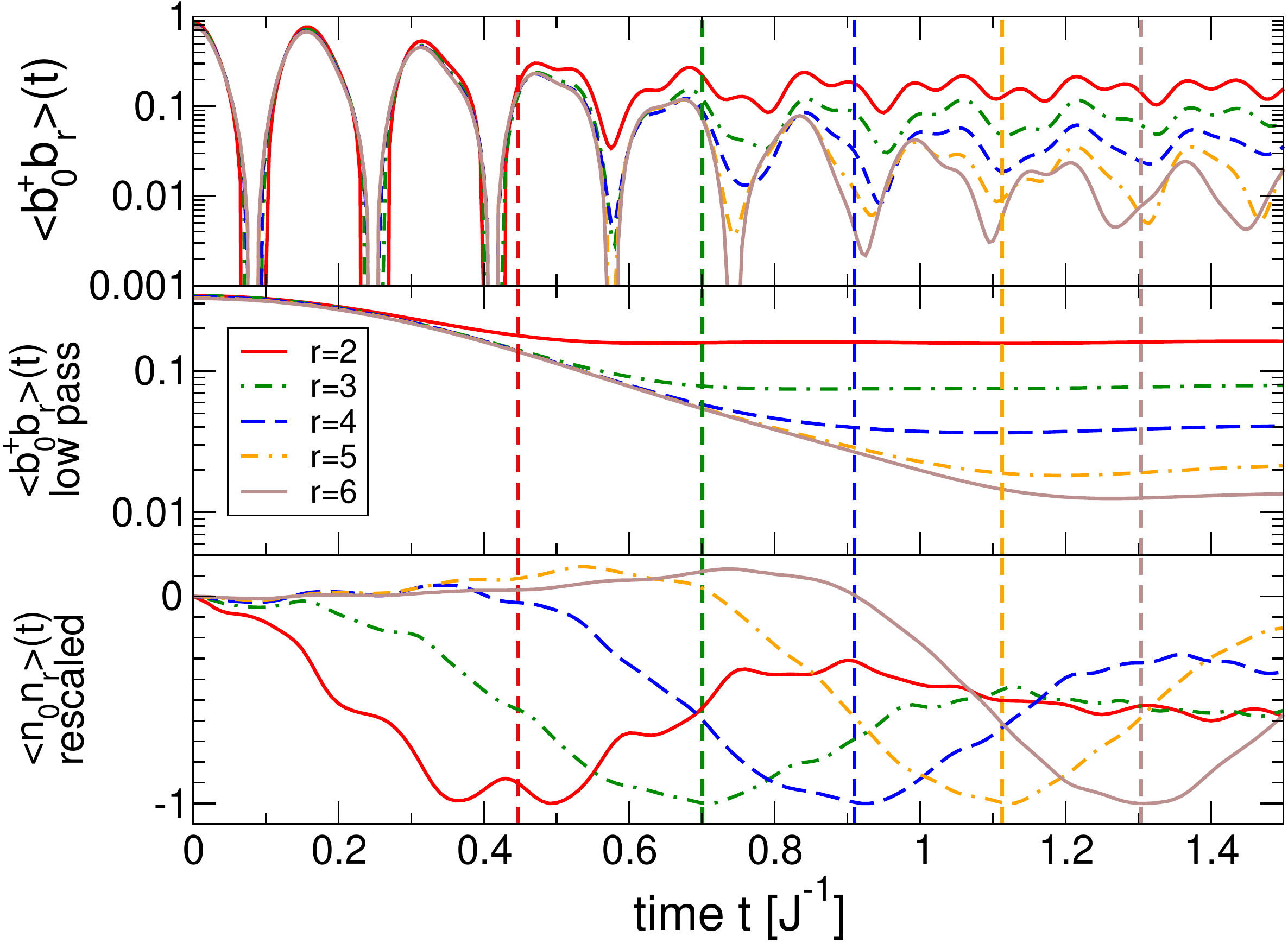}
\end{indented}
 \caption{Time-evolution of correlation functions after a quench from $U_i=2J$ to $U_f=40J$. The upper panel shows the single particle correlation functions
     $\aver{b^\dagger_0 b_{r}}$  
     for different distances $r$. The correlations show partial revivals
     up to a time $t_r$ when they start to reach a quasi-steady state. This time
     $t_r$ grows approximately linearly with the distance $r$ as marked by the vertical 
     lines. The central panel shows the same correlations functions after filtering out the high frequencies, see text for details. 
     The lowest panel shows the density density correlations function  $\aver{n_0 n_{r}}$ after shifting and rescaling their amplitude for better visibility. 
     The common vertical dashed lines denote the arrival of the minima as determined from the density-density correlations. 
     The data shown is ED for a $L=14$ and DMRG data for $L=32$ and filling $n=1$.      }   
 \label{fig:outcoupling}
 \end{figure} 
\begin{figure}[ht]
\begin{indented}
 \item\includegraphics[width=0.8\linewidth]{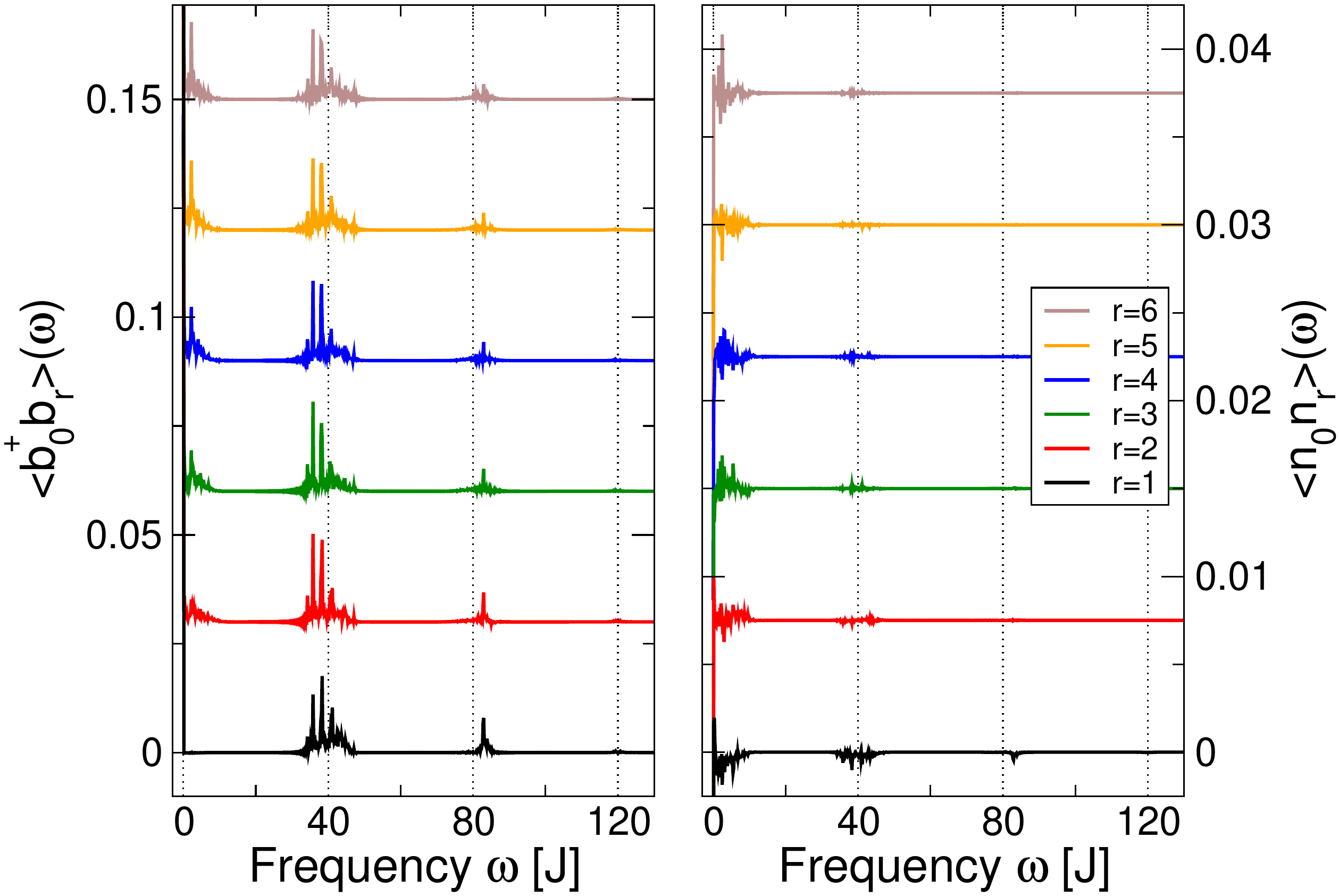}
\end{indented}
 \caption{Fourier transform of the equal-time single-particle and density-density correlation functions for different distances $r$. The same parameters as in \Fref{fig:outcoupling} are used. 
 In the single-particle correlation function clear frequency bands located at multiples of the interaction strength $U_f=40J$ can be seen. 
 The density-density correlation is dominated by the contributions at low frequencies from zero up to $\sim J$. } 
  \label{fig:shorttime}
\end{figure}

In this section we investigate how correlations over a distance $r$ react to the sudden parameter change. We consider two different types of correlations, the single particle correlations $\aver{b^\dagger_j b^{\phantom{\dagger}}_{j+r}}$ and the density-density correlations $\aver{n_jn_{j+r}}$ at equal time.
In Fig.~\ref{fig:outcoupling} we show the time-evolution of the different correlations after a quench from the superfluid, $U_i=2$, to the Mott-insulating, $U_f=40$, parameter regime. 

\paragraph{Single-particle correlations}
The upper panel shows the correlations $\aver{b^\dagger_0
  b^{\phantom{\dagger}}_{r}}$ for different distances $r$\footnote{To extract these correlations from the DMRG data with open boundary conditions the average over central sites is taken. Note that for periodic boundary conditions this quantitiy is real due to symmetry, whereas for open boundary conditions an imaginary part can develop. However for the shown functions and times the imaginary part is negligible.  }. For short times the single particle correlations oscillate with the period $2\pi/U_f$. The origin of these oscillations lies in the integer spectrum of the operator $\hat{n}_j(\hat{n}_j-1)/2$. Consider the limit of very strong interactions, where the
time-evolution is totally dominated by the interactions. The time evolution of the single particle correlations is given by  
$$\aver{b^\dagger_i b^{\phantom{\dagger}}_{j}}(t) =
\sum_{\{m\},\{m'\}} \delta_{m_i,m'_i+1} \times \delta_{m_j,m'_j-1} \times
  \e^{\textrm{i} U_f (m'_j-m'_i-1)t} c^*_m c_{m'} \bra{\{m \}} b^\dagger_i b_j
  \ket{\{m' \} }. $$ 
Here we use the notation $\{m\}$ for the Fock state with $m_i$ particles on site $i$. The time-evolution
of the correlation function is determined by the non-vanishing cross terms $\bra{\{m \}} b^\dagger_i b_j \ket{\{m' \}}$ of Fock states
whose occupations vary by removing one particle from site $j$ and adding it at site $i$. The frequencies occuring in the time-evolution will be given by $U_f(m'_j-m'_i-1)$. This results in oscillations of period $T\approx 2\pi/U_f$ for equally occupied lattice sites $j$ and $i$ of the state $\ket{\{m'\}}$. For unequal onsite occupations higher multiples of $U_f$ can occur in the frequency spectrum.

Thus the distribution of the different frequencies in the Fourier transformation of the single particle correlation functions (left panel in Fig.~\ref{fig:shorttime}) and the occupation difference in the initial state are intimately connected.
For the shown quench ($U_i=2J, U_f=40J$) sizable contributions of the lowest three
frequency bands can be seen, whereas the occupation of higher frequency bands
becomes very small. The distribution among the frequency bands changes by varying the initial value of the interaction
strength $U_i$.  If the initial state is deep in the superfluid regime, the peaks at higher frequencies show more weight due to the presence of strong particle number fluctuations. In contrast if the initial state is close to the Mott-insulating transition the particle fluctuations are suppressed and the weight of the peaks at higher frequencies decreases accordingly. The width of the frequency bands is due to the finite value of $J$ and is responsible for the decay of the oscillations in time \cite{KollathAltman2007}.

Let us now come back to the real-time evolution of the single particle correlations for different distances.
After a time $t\approx 0.5/J$ the correlation corresponding to the smallest distance shown, $r=2$, reaches a quasi-steady value (marked by the leftmost vertical line). The correlations of distance $r=3$ deviate from the correlation with $r>3$ for a time $t\approx 0.7/J$ (second left vertical line). The same can be observed for correlations with increasing
distances at longer times. In the central part of the
\Fref{fig:outcoupling} the same single particle correlations are shown, but the high frequency oscillations at frequencies $\omega_n \sim n \times U_f $ with $n\gtrsim 1$ are filtered out, and only the $n=0$ contributions are kept. In these low-pass filtered correlations the saturation to a quasi-steady state value
can be seen much more clearly. 

\begin{figure} [ht]
\begin{indented}
\item\includegraphics[width=0.8\linewidth]{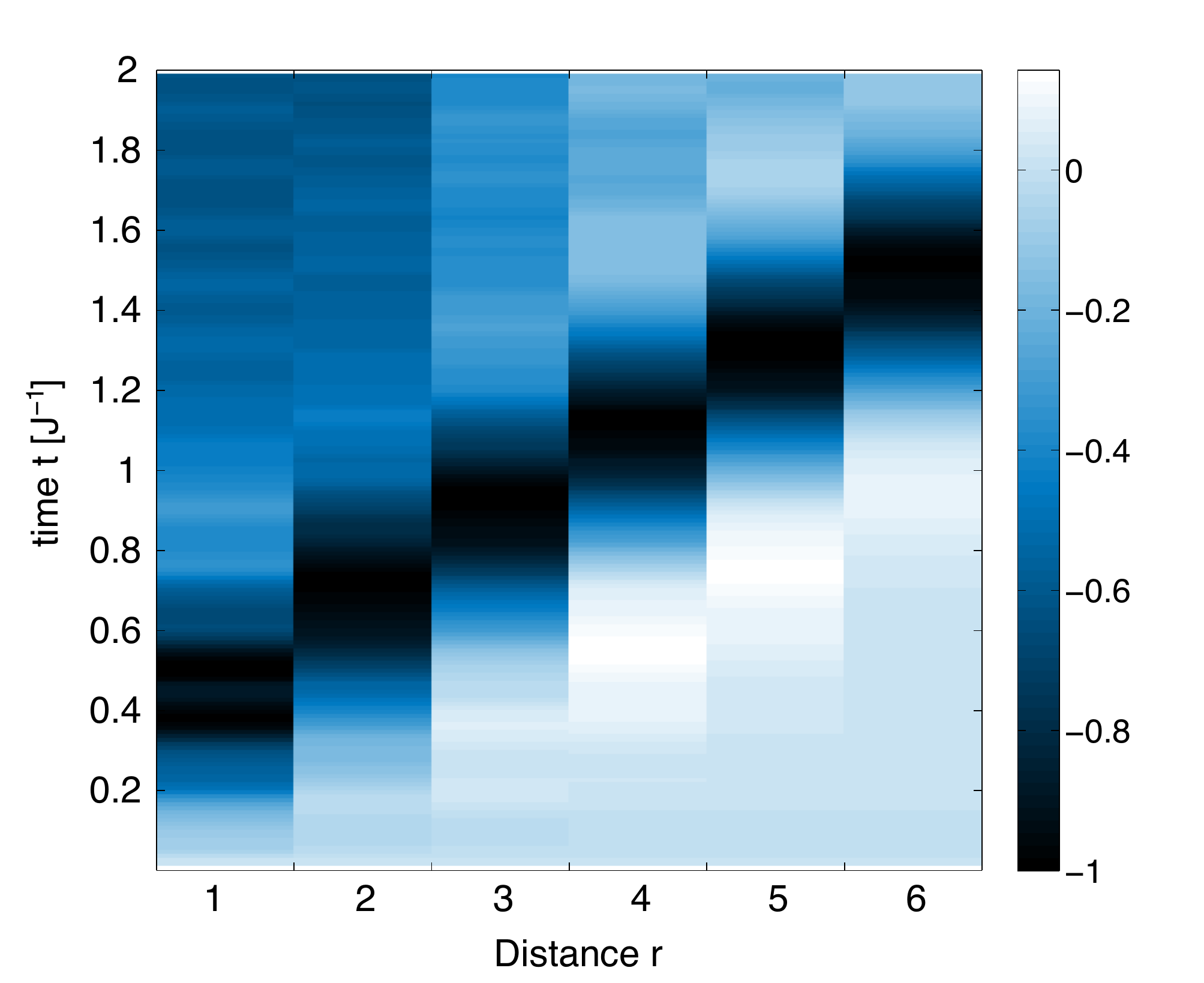}
\end{indented}
\caption{ 
  Time-evolution of the rescaled density-density correlations $\aver{n_0 n_{r}}(t)$
  after a quench with the same parameters as in \Fref{fig:outcoupling}. The front of the evolution evolves
  approximately with a constant velocity.
  }
  \label{fig:densitydensity}
\end{figure} 

\paragraph{Density-density correlations}
The lowest panel of \Fref{fig:outcoupling} shows the density-density correlations $\aver{n_jn_{j+r}}-\aver{n_j}\aver{n_{j+r}}$ at equal time. The amplitudes of the correlations are rescaled and shifted for better readability \footnote{We will denote the rescaled function $\aver{n_jn_{j+r}}-\aver{n_j}\aver{n_{j+r}}$ averaged over different sites $j$ in the center of the chain by $\aver{n_0n_{r}}$.}. 

The density-density correlations do not show strong oscillations, but remain almost constant in time up to the moment, where a pronounced signal arrives.
In their Fourier spectrum (right panel of \Fref{fig:shorttime}) mostly low frequencies $\sim J$ occur. This is due to the fact that the interaction term of the Hamiltonian commutes with the correlation function. Thus, in the strong coupling limit the interaction term does not give rise to oscillations. Approximately at the same time as the single particle correlations saturate, the spreading of a signal (here the reaching of a minimum, loci of the common vertical dashed lines in \Fref{fig:outcoupling}) can be found in the density-density correlation $\aver{n_0 n_{r}}$.

In \Fref{fig:densitydensity} we show a contour plot of the rescaled density correlations. In this representation a clear light cone evolution can be seen, i.e.~a front travels through the system at almost constant speed. 
Let us note, that the same light cone effect occurs in the evolution of the correlations for different quench parameters (cf.~\Tref{tab:vel}) and also in incommensurate systems, e.g.~at filling $n=1/2$. 
 
\begin{figure} [ht]
\begin{indented}
\item
\item\includegraphics[width=0.8\linewidth]{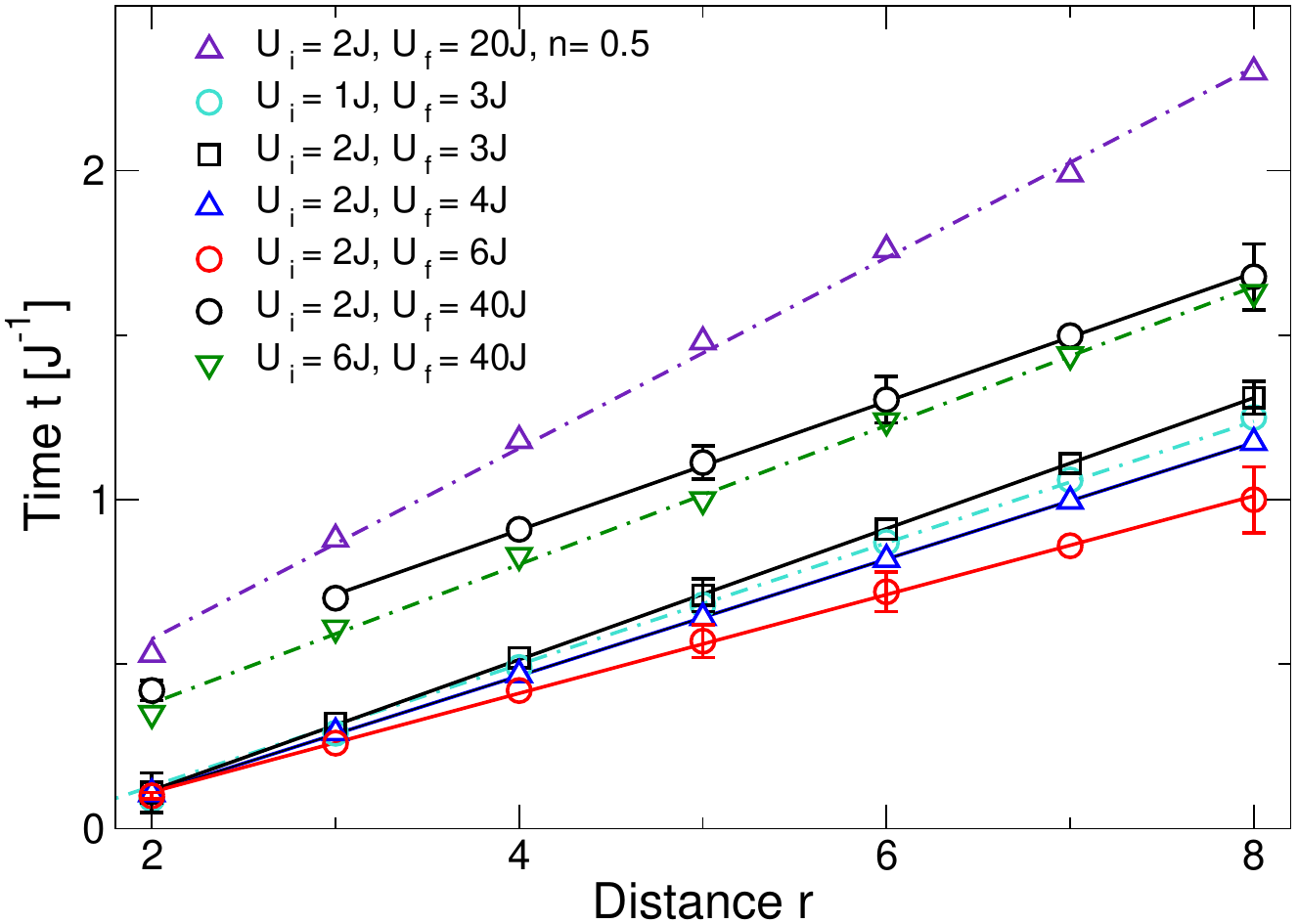}
\end{indented}
  \caption{Characteristic time $t_r$ for arrival of the signal in the density-density correlations depending on the distance $r$ are summarized for different quench parameter $U_i$, $U_f$, and average density $n$. A clear linear relation between time and distance is observed for all shown quenches. To give an order of the uncertainties, typical error bars are plotted for chosen points. They take into account both the difficulty to identify a sharp signal and deviations between different system length ($L=14$ up to $L=64$).
}  
\label{fig:velocity}
\end{figure} 
\begin{figure} [ht]
\begin{indented}
\item
\centerline{\includegraphics[width=0.8\linewidth]{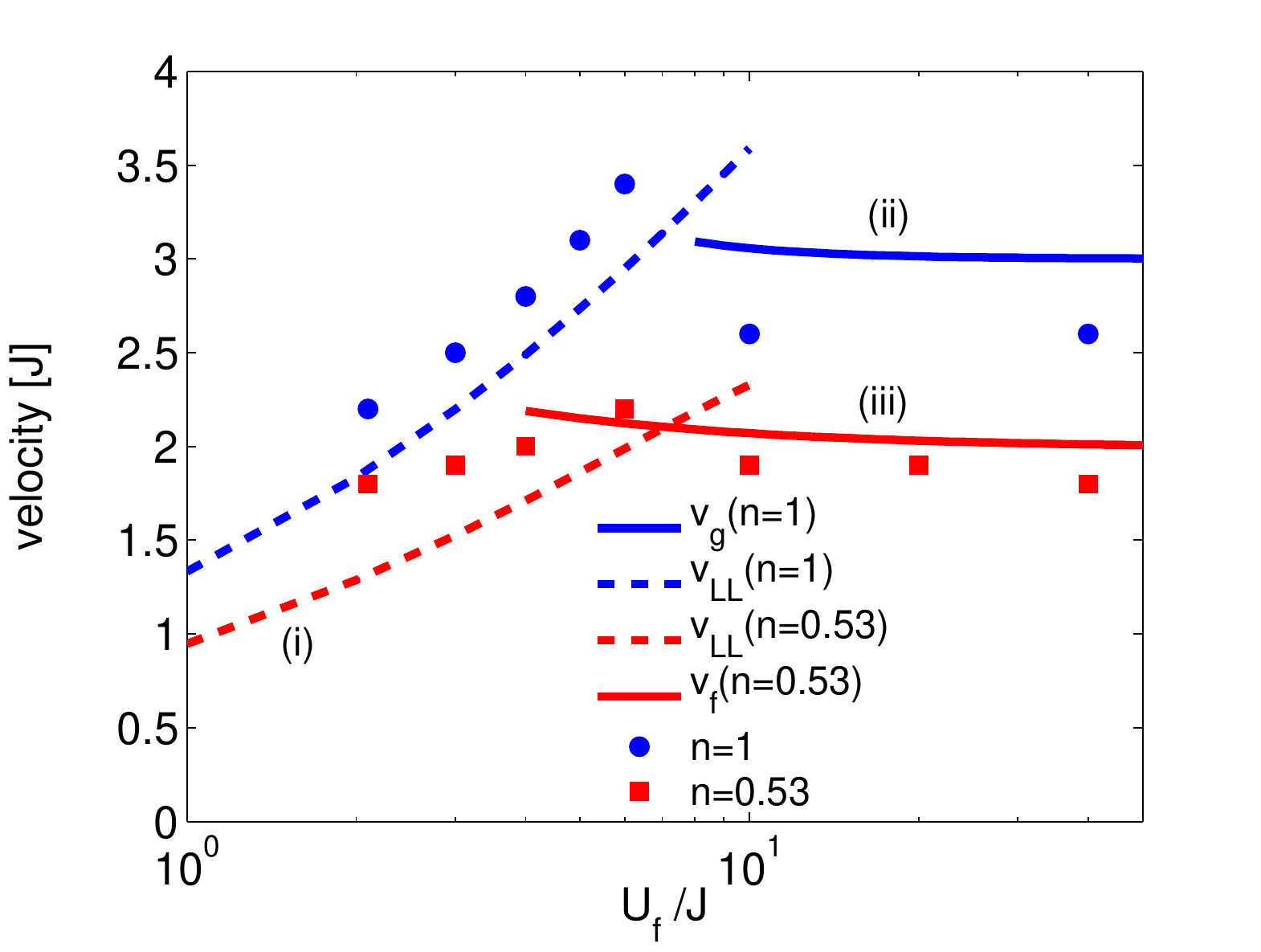}}
\end{indented}
  \caption{Dependence of the velocity $v_s/2$ on the final interaction strength after the quench. Results are shown for different values of the density and compared to an analytical approximation described in the text. The initial value is $U_i=2J$.}
\label{fig:vel}
\end{figure} 
 
\paragraph{Propagation velocity $v_s$}
To extract the velocity $v_s$ of the signal propagation in the density-density correlations, we plot in \Fref{fig:velocity} the time $t_r$, when a signal, e.g. the minimum, occurs in the density-density correlations versus the distance $r$. The curves for different values of the initial and final interaction strength $U_i$ and $U_f$ are presented. A clear linear behaviour of the time $t_r$ versus the distance is seen. The shift between the curves, e.g.~for $U_f=20,40$ and $U_f=4,6$, stems from the different signatures that have been tracked in the density-density correlations. The inverse of the slope is the signal propagation velocity $v_s$ of the signature. We extract the velocity $v_s$ by a linear fit $t_r=r/v_s+b$, where $v_s$ and $b$ are the fitting parameters. %
The results for $v_s$ are shown in \Tref{tab:vel}. 

\begin{table}
\caption{Velocity extracted from a linear fit.
} 
\begin{indented}
\item[]
\begin{multicols}{2}
\begin{minipage}{0.8\linewidth}
\begin{tabular}{| l | l | l |l |}
\hline
$n$ & $U_i/J$ & $U_f/J$ & $v_s [J] $ \\
\hline 
\hline
1 & 2 & 2.1& 4.4\\
1 & 2 & 3& 5\\
1 & 2& 4 & $5.6\pm0.4$\\
1 & 2 & 5 & $6.2$\\
1 & 2& 6 &6.8 \\
1 & 2& 10 &5.2 \\
1 & 2& 40 & $5.2 \pm 0.6$\\
1 & 1&4& 5.9\\
1 & 3&4& 5.6\\
1 & 6 & 40 & 4.8 \\
1 & 1& 3& 5.2\\
\hline
\hline
\end{tabular}
\end{minipage}
\begin{minipage}{0.5\linewidth}
\begin{tabular}{| l | l | l |l |}
\hline
$n$ & $U_i/J$ & $U_f/J$ & $v_s [J] $ \\
\hline 
\hline
0.5 & 2& 2.1 & 3.6 \\
0.5 & 2& 3 & $3.8$\\
0.5 & 2 & 4 & 4 \\
0.5 & 2 & 6 & 4.4 \\
0.5 & 2& 10 & 3.8 \\
0.5 & 2& 20 & $3.8$\\
0.5 & 2 & 40 & 3.6 \\
\hline
\hline
\end{tabular}
\end{minipage}
\end{multicols}
\end{indented}
\label{tab:vel}
\end{table}

The velocity of the spreading of correlations after a quench has been identified by Calabrese and Cardy \cite{CalabreseCardy2006} within conformal field theory and for different integrable models to be twice the maximal mode velocity. The simple picture given is that the modes depart from both considered sites and the signal arrives, if both modes interfere. In the description by a conformal field theory this velocity agrees with the sound velocity of the system, since the dispersion relation is linear. In a chain of harmonic oscillators or in a spin chain lattice effects occur, which cause a curved dispersion relation. Thus the maximal velocity of a mode can be distinct from the sound velocity in the system. For the transverse Ising model a velocity equal to one has been found \cite{IgloiRieger2000}.

In \Fref{fig:vel} we show the results for the rescaled signal velocity $v_s/2$ for one initial interaction strength $U_i=2J$. We present results for different densities. At the commensurate density $n=1$ a phase transition to the Mott-insulator takes place at a critical value. In contrast for the incommensurate density $n=1/2$ \footnote{In the open $L=32$ system used in the t-DMRG, the density in the middle of the system is $n\approx 0.53$.} the system stays in the superfluid phase for all interaction strengths. The signal velocity shows a strong increase for low interaction strength. After reaching a maximum around $U/J \approx 6$ it saturates for strong interactions to an almost constant value. 

We further compare our results to different characteristic velocities of the Bose-Hubbard model: (i) the sound velocity in the superfluid regime, (ii) the maximal mode velocity of a simplified model in the Mott-insulator, (iii) the maximal mode velocity of a fermionic model applicable at low filling.

(i) For an infinitesimal quench inside the superfluid phase, the rescaled signal velocity $v_s/2$ for the spreading of the correlations can be described by the sound velocity of the system. However, for finite quenches we expect the rescaled signal velocity to be larger than the sound velocity, since modes with higher velocities can be excited. In \Fref{fig:vel} we approximate the sound velocity in the superfluid regime by the sound velocity of the corresponding continuous Lieb-Liniger model \cite{Lieb1963}. For small values of $\gamma$, it is given by $v_{LL}=2 n \sqrt{\gamma}\sqrt{1-\frac{\sqrt{\gamma}}{2\pi}}$, where $\gamma=U/(2Jn)$. This expression approximates the sound velocity for the Bose-Hubbard model up to an interaction strength $\gamma \lesssim 4$ \cite{KollathZwerger2004}. 

(ii) 
An idea of the expected signal velocity in the Bose-Hubbard model in the Mott-insulating state can be obtained by mapping the system onto a simpler model using only three local states, e.g.~occupation by $n_0-1$, $n_0$, and $n_0+1$ bosons per site, where $n_0$ is an integer \cite{AltmanAuerbach2002,HuberBlatter2007}. In the Mott-insulating phase the dispersion relation for a particle hole excitations in this effective model can be determined as $\epsilon(k)=\sqrt{U^2-U \epsilon_0(k) (4n_0+2)+\epsilon_0(k)^2}/2$ \cite{HuberBlatter2007}. Here $\epsilon_0(k)=2J \cos(ka)$ is the band dispersion for the non-interacting case. The group velocity in this case is given by $ v=\frac{\partial \epsilon(k)}{\partial k} $. The maximum velocity $v_g$ of a mode given by this model is shown in \Fref{fig:vel}. In the strong coupling limit this velocity agrees with the velocity extracted from a perturbative calculation where the maximal group velocity is given by 
$aJ/\hbar (2n_0+1)$. For the case of $n=1$ this results in $3aJ/\hbar$. In particular we see that towards the critical value in this model the velocity slightly increases. The momentum at which the maximum velocity is reached changes compared to the strong coupling limit. 

(iii) For strong interaction and low filling the Bose-Hubbard system can be mapped onto a fermionic system \cite{Cazalilla2004}. In this fermionic system the velocity is given by $v_f=2\sin(\pi n) (1-8J n\cos(\pi n)/U)$.

In \Fref{fig:vel} we compare our numerical results to the different velocities (i)-(iii). For small interaction strength we see that the velocity $v_s/2$ is always larger than the sound velocity showing a similar rise with $U/J$. Comparing further the velocity of different quenches in the superfluid regime, e.g.~$U_i=1,2$ and $U_f=4$ in  \Tref{tab:vel}, the velocity $v_s/2$ seems to approach the sound velocity if the parameter changes becomes smaller.

In the regime of strong interaction the qualitative features of the signal velocity are well reproduced by the given approximations. In particular, the velocity is almost constant for high interaction values and increases if $U/J$ is lowered towards $U/J\approx 6$. Further the order of magnitude of the values is in good agreement. 

However, the approximation has to be taken with care. It does not take all higher energy excitations into account, which will be neccessary to quantitatively describe the situation under consideration. 
 Further it does not take into account the initial state, i.e.~which of the modes are actually excited by the quench. This is in contrast to the numerical results in \Tref{tab:vel} which suggest that the value of the observed signal velocity might depend as well on the initial state. In particular if the higher particle fluctuations are present in the initial state the observed signal velocity seems to be larger. 

\section{Dynamics of entanglement and mutual information}

We turn in the following to the time-evolution of the entanglement and the amount of correlations 
between different subsystems after a quench. As a measure for the entanglement we use the von Neumann block entropy 
(see section \ref{sec:static}) and the mutual information. Whereas the von Neumann entropy describes the entanglement of a region of the system with the remaining part, the mutual information gives a measure about the amount of correlations between different subsystems~\cite{GroismanWinter2005}. In the following we first analyze the time-evolution of the von Neumann entropy before we turn to the evolution of the mutual information. 

\paragraph{von Neumann entropy}
\begin{figure} [ht]
\begin{indented}
\item\includegraphics[width=0.8\linewidth]{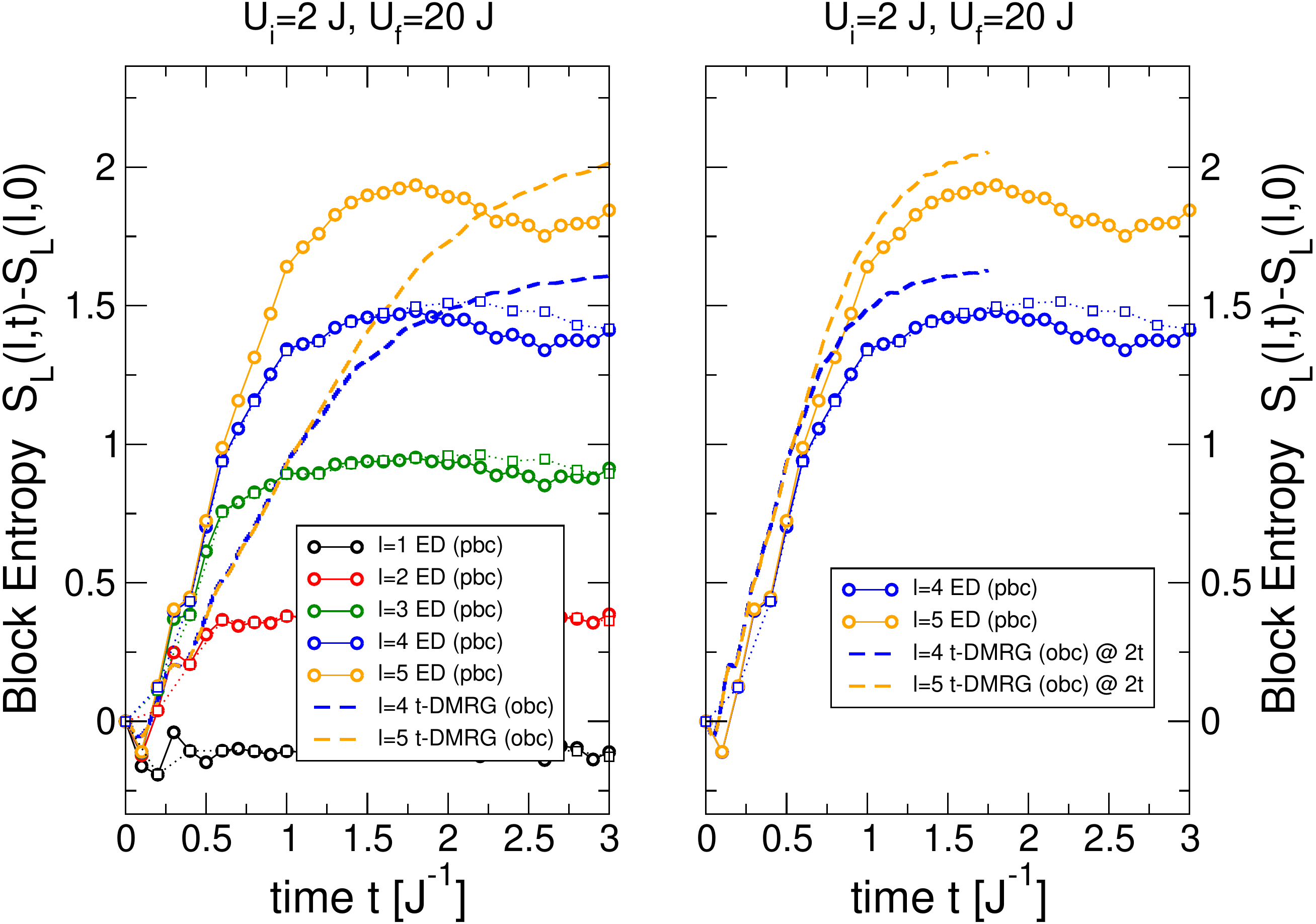}
\end{indented}
\caption{ 
  Time-evolution of the block entropy for different block lengths for a quench
  from $U_i=2J$ to $U_f=20 J$. 
  The initial linear rise is the same for different block lengths, but depends on the boundary condition of the block
  (PBC: two interface links per block, faster rise of the entropy; OBC:~one interface link per block, slower rise of the entropy).
  For longer times a saturation of the entropy depending linearly on the block length is observed.
  The left panel contains the original time evolution, while in the right panel, the obc data has been plotted for
  times $\tilde{t}=2t$, so as to illustrate the slower increase of the entropy due to the smaller boundary.
  The PBC results are from ED on $L=14$ (circles) and $L=16$ (squares) systems, while the OBC data has been obtained
  using $t-$DMRG up to $L=64$.
}
\label{fig:entropy}
\end{figure} 
In Fig.~\ref{fig:entropy} we show the time-evolution of the von Neumann entropy for Bose-Hubbard systems with periodic (ED, $L=14,16$
\footnote{Due to the computational challenge of calculating density matrices in ED for large blocks we chose to limit the local boson
occupancy to 3 at most.})
and open boundary conditions (t-DMRG, 5 bosons per site are allowed). In both cases we focus on contiguous blocks of length~$l$. 
For open boundary conditions the blocks are aligned with one of the boundaries. We plot the difference
$S_L(l,t)-S_L(l,0)$ so that the curves for all block sizes start at zero at $t=0$.

In the left panel we show data for ED ($l=1,2,3,4,5$) and DMRG ($l=4,5$) results for a specific quench from
$U_i=2 J$ to $U_f=20 J$.
Let us first discuss the ED results: 
Immediately after the quench for $t\lesssim 0.4 J^{-1}$ a small dip shows up in all block entropies. However,
from time $t\approx 0.4 \hbar /J$ up to $t^*(l)$ a linear rise of the
block entropies can be observed. Interestingly all block sizes for $l>1$
have the same slope, until they bend over to an almost flat behaviour at 
successively later times. The dip in the entropy at larger times is a finite size effect, as can
be seen by comparing the data for different system sizes (circles for $L=14$  and squares for $L=16$).
The saturation value of the different block entropies
depends linearly on the block size, defining an entropy propagation velocity $v_\mathrm{e}$,
which is roughly equal to $v_\mathrm{s}$ determined based on the density-density correlation 
functions in the preceding~\sref{sec:light_cone}.
In a next step it is instructive to compare the entanglement dynamics between different block geometries 
(two interface links in ED, one interface link in $t-$DMRG). The $t-$DMRG results 
are also shown in the left panel of \Fref{fig:entropy}. The entropy of these open blocks increases more slowly,
but converges to about the same value at late times as the periodic blocks of the same length. To
illustrate this convincingly we show in the right panel the same data, but where the entropy of the open
blocks is shown on a time scale which is twice as fast. Indeed the results of the two block geometries agree reasonably well. This nice result lends direct support to the picture developed by Calabrese and Cardy~\cite{CalabreseCardy2005},
where they predicted that the saturation of the entropy occurs at $t^*_\mathrm{PBC}(l) =l/2v$ for periodic boundary conditions, while
$t^*_\mathrm{OBC}=l/v$ for blocks aligned with an open boundary. 

These characteristics of the entanglement evolution are similar to
the results obtained for different models in Refs.~\cite{CalabreseCardy2005,AmicoMassimapalma2004,ChiaraFazio2006,EisertOsborne2006, AmicoVedral2008, BarmettlerGritsev2008}. A linear growth of the entropy has been seen up to times $t=l/2v$, where $v$ is
the maximal velocity of the excitations. 
Afterwards a saturation of the entropy is seen for $t\to \infty$, and the rate of the approach to the saturation value 
is related to the dispersion relation of the underlying model.
At a fixed time the entanglement saturates for increasing block length, i.e.~fullfils a boundary law with the boundary
increasing with time. This shows that the boundary law for the dynamics of
entanglement which has been proven mostly for 1D spin-systems~\cite{EisertOsborne2006,BravyiVerstraete2006}
seems to be valid in more general systems such as the Bose-Hubbard system considered here. 

\paragraph{Mutual information}

\begin{figure} [ht]
\begin{indented}
\item\includegraphics[width=0.8\linewidth]{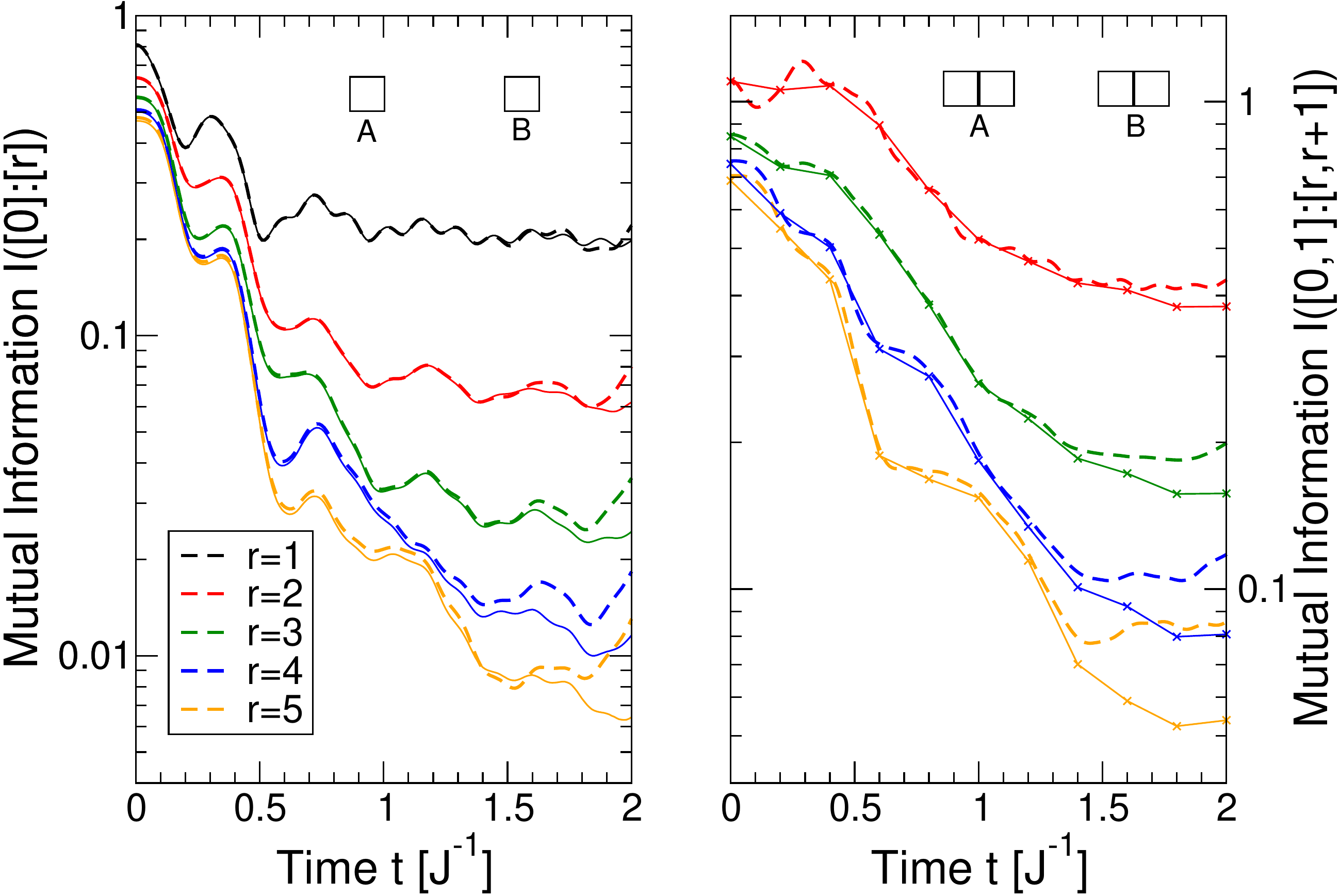}
\end{indented}
\caption{(Color online) 
  Time-evolution of the mutual information \protect{\eref{eqn:mutualinfo}} between two blocks of equal length $l=1$ 
  (left panel) and $l=2$ (right panel) for different spatial
  separations $r$. The quench parameters are $U_i=2J$, $U_f=20J$, and $n=1$. 
  The results have been obtained by ED for $L=14$ (bold, dashed lines) and $L=16$ (thin, straight lines).
}
\label{fig:mutual}
\end{figure} 
While the von Neumann entropy is rapidly increasing in time and finally leads to an extensive entropy scaling with the block volume at large times,
it is interesting to ask whether the vast entanglement entropy also leads to increased correlations between subparts of the system. A very 
useful quantity to address this question is the mutual information. The mutual information $I(A:B)$ between two subsystems A and B of the system is defined by 
\begin{equation}
I(A:B)= S_A+S_B-S_{A \cup B}.
\label{eqn:mutualinfo}
\end{equation} 
 Here $S_X$ is the von Neumann entropy for the subsystem $X$. The mutual information $I(A:B)$ measures the total amount of information of system $A$ about system $B$~\cite{GroismanWinter2005}. Note that $A \cup B$ is not required to be equal to the total system.
 Interestingly the mutual information can fulfill an area law at finite temperatures, while the entropy is expected to follow a volume law~\cite{WolfCirac2008}~\footnote{In this reference the authors study the case of the total system equal to $A\cup B$.}.

In \Fref{fig:mutual} we show the time evolution of the mutual information between two blocks of $l=1$ (left panel) or $l=2$ (right panel) sites each, 
shifted by a distance $r$. 
At $t=0$ we expect the mutual information to decay slowly with the block separation $r$, 
since the starting state is basically a scale invariant critical state and the
mutual information is just a complicated function of the basic critical correlation functions, such as $\langle b^\dagger_0 b_{r}\rangle$ and 
$\langle n_0 n_{r}\rangle$, which all decay algebraically. As a function of time $t$, the mutual information is decreasing rapidly with some 
slow oscillations on top. For later times it seems as though the mutual information levels off
to a finite value at a time  
which depends again linearly on the distance $r$. The mutual information at late times 
decays much faster as a function of distance than in the initial state. So even though the entanglement entropy is vastly growing in time, this does not
lead to enhanced entanglement between different subsystems.

\section{Conclusion}

In our work we show that correlations and entanglement are very useful quantities to characterize the equilibrium and dynamic properties of a quantum many body system. 

In the first part of our work we showed that the static von Neumann entanglement signals the quantum phase transition between the superfluid and Mott-insulating state without previous knowledge on the type of the phase transition. In the superfluid, the von Neuman entropy is in very good agreement with previous predictions by conformal field theory \cite{HolzeyWilczek1994,CalabreseCardy2004} with a central charge $c=1$. Deviation are only found close to the boundaries of the system. These deviations are induced by the inhomogeneous density distribution caused by the open boundary conditions. A saturation of the entropy with the block length is found in the gapped Mott-insulating phase as predicted \cite{VidalKitaev2003,CalabreseCardy2004}.

In the second part of our work, the time-evolution after a sudden parameter change is analyzed with a focus on the spreading of information. Hereby different parameter changes are discussed ranging from the change between the superfluid to Mott-insulating phase over a quench inside the superfluid to a quench inside the Mott-insulating regime. Our study proposes that the Lieb Robinson theorem is valid as well in the considered situation of the Bose-Hubbard model. This relies on our findings that a light-cone like evolution takes place in different correlation functions. The velocity of the front evolving in the correlation functions is discussed and compared to different characteristic velocities of the system. The validity of the Lieb-Robinson theorem is further supported by the von Neumann entropy which shows for a certain block length a linear growth for short times. To a good approximation the rate of the 
block entropy increase is shown to follow a boundary law, while the entropy value at saturation 
seems only to depend on the block volume. Our results nicely corroborate recent analytical 
results \cite{CalabreseCardy2005,EisertOsborne2006,BravyiVerstraete2006}.
However, in contrast to the von Neumann entropy after a quench from the superfluid to the Mott-insulating regime, the mutual information between two  spatially separated small blocks relaxes to a lower value than in the starting state. 
So even though the entanglement entropy is vastly growing in time, this does not necessarily lead to enhanced entanglement between different regions of the system.

\section{Acknowledgement}
We would like to thank E. Altman, J. Eisert, S. Huber, S. Manmana, A. Muramatsu, R. Noack, A. Rosch, and S. Wessel for fruitful discussions. This
work was partly supported by the Swiss National Science Foundation.
We wish to thank the Institute Henri Poincare-Centre Emile Borel for hospitality and support. Further CK acknowledges support by the RTRA network 'Triangle de la Physique' and the DARPA OLE program.

The ED simulations have been performed on the machines of the CSCS (Manno). 


\section*{References}
\bibliographystyle{unsrt}
\bibliography{references130308}

\begin{thebibliography}{10}

\bibitem{AmicoVedral2008}
Luigi Amico, Rosario Fazio, Andreas Osterloh, and Vlatko Vedral.
\newblock Entanglement in many-body systems.
\newblock {\em Rev. Mod. Phys.}, 2008.

\bibitem{BlochZwerger2007}
I.~Bloch, J.~Dalibard, and W.~Zwerger.
\newblock Many-body physics with ultracold gases.
\newblock {\em arXiv:0704.3011}, 2007.

\bibitem{LiebRobinson1972}
E.~H. Lieb and D.~W. Robinson.
\newblock The finite group velocity of quantum spin systems.
\newblock {\em Communications in Mathematical Physics}, 28:251--257, 1972.

\bibitem{CramerEisert2008}
M.~Cramer, A.~Serafini, and J.~Eisert.
\newblock Locality of dynamics in general harmonic quantum systems.
\newblock {\em arXiv:0803.0890}, 2008.

\bibitem{IgloiRieger2000}
F.~Igl\'oi and H.~Rieger.
\newblock Long-range correlations in the nonequilibrium quantum relaxation of a
  spin chain.
\newblock {\em Phys.~ Rev.~ Lett.}, 85:3233, 2000.

\bibitem{RigolOlshanii2006a}
Marcos Rigol, Alejandro Muramatsu, and Maxim Olshanii.
\newblock Hard-core bosons on optical superlattices: Dynamics and relaxation in
  the superfluid and insulating regimes.
\newblock {\em Physical Review A (Atomic, Molecular, and Optical Physics)},
  74(5):053616, 2006.

\bibitem{Cazalilla2006}
M.~A. Cazalilla.
\newblock Effect of suddenly turning on interactions in the luttinger model.
\newblock {\em Physical Review Letters}, 97(15):156403, 2006.

\bibitem{CubittCirac2007}
T.S. Cubitt and J.I. Cirac.
\newblock Engineering correlation and entanglement dynamics in spin systems.
\newblock {\em quant-ph/0701053}, 2007.

\bibitem{CramerOsborne2008}
M.~Cramer, C.~M. Dawson, J.~Eisert, and T.~J. Osborne.
\newblock Exact relaxation in a class of nonequilibrium quantum lattice
  systems.
\newblock {\em Physical Review Letters}, 100(3):030602, 2008.

\bibitem{CalabreseCardy2006}
Pasquale Calabrese and John Cardy.
\newblock Time dependence of correlation functions following a quantum quench.
\newblock {\em Phys.~ Rev.~ Lett.}, 96(13):136801, 2006.

\bibitem{CalabreseCardy2007}
Pasquale Calabrese and John Cardy.
\newblock Quantum quenches in extended systems.
\newblock {\em J.Stat.Mech.}, P008, 2007.

\bibitem{GreinerBloch2002}
M.~Greiner, O.~Mandel, T.~Esslinger, T.~W. H\"ansch, and I.~Bloch.
\newblock Quantum phase transition from a superfluid to a {M}ott insulator in a
  gas of ultracold atoms.
\newblock {\em Nature}, 415:39, 2002.

\bibitem{HolzeyWilczek1994}
C.~Holzey, F.~Larsen, and F.~Wilczek.
\newblock Geometric and renormalized entropy in conformal field theory.
\newblock {\em Nucl. Phys. B}, 424:44, 1994.

\bibitem{CalabreseCardy2004}
Pasquale Calabrese and John Cardy.
\newblock Entanglement entropy and quantum field theory.
\newblock {\em J.~Stat.~Mech.: Theor.~Exp.~}, P06002, 2004.

\bibitem{KollathAltman2007}
C.~Kollath, A.M. L\"auchli, and E.~Altman.
\newblock Quench dynamics and non-equilibrium phase diagram of the
  {B}ose-{H}ubbard model.
\newblock {\em Phys.~ Rev.~ Lett.}, 98:180601, 2007.

\bibitem{CalabreseCardy2005}
Pasquale Calabrese and John Cardy.
\newblock Evolution of entanglement entropy in one-dimensional systems.
\newblock {\em J.~Stat.~Mech.: Theor.~Exp.~}, P04010, 2005.

\bibitem{EisertOsborne2006}
Jens Eisert and Tobias~J. Osborne.
\newblock General entanglement scaling laws from time evolution.
\newblock {\em Phys. Rev. Lett.}, 97(15):150404, Oct 2006.

\bibitem{BravyiVerstraete2006}
S.~Bravyi, M.~B. Hastings, and F.~Verstraete.
\newblock Lieb-{R}obinson bounds and the generation of correlations and
  topological quantum order.
\newblock {\em Phys.~ Rev.~ Lett.}, 97(5):050401, 2006.

\bibitem{JakschZoller1998}
D.~Jaksch, C.~Bruder, I.~Cirac, C.~W. Gardiner, and P.~Zoller.
\newblock Cold bosonic atoms in optical lattices.
\newblock {\em Phys.~ Rev.~ Lett.}, 81(15):3108, 1998.

\bibitem{Zwerger2003}
W.~Zwerger.
\newblock {M}ott-{H}ubbard transition of cold atoms in optical lattices.
\newblock {\em Journal of Optics B:}, 5:9, 2003.

\bibitem{FisherFisher1989}
M.~P.~A. Fisher, P.~B. Weichman, G.~Grinstein, and D.~S. Fisher.
\newblock Boson localization and the superfluid-insulator transition.
\newblock {\em Phys.~Rev.~B}, 40(1):546, 1989.

\bibitem{White1992}
S.~R. White.
\newblock Density matrix formulation for quantum renormalization groups.
\newblock {\em Phys.~ Rev.~ Lett.}, 69:2863, 1992.

\bibitem{Hallberg2006}
K.~Hallberg.
\newblock New trends in density matrix renormalization.
\newblock {\em Adv.~Phys.~}, 55:477, 2006.

\bibitem{Schollwoeck2005}
U.~Schollwock.
\newblock The density-matrix renormalization group.
\newblock {\em Reviews of Modern Physics}, 77(1):259, 2005.

\bibitem{park:5870}
Tae~Jun Park and J.~C. Light.
\newblock Unitary quantum time evolution by iterative lanczos reduction.
\newblock {\em J.~ Chem.~ Phys.}, 85(10):5870--5876, 1986.

\bibitem{manmana-2005-789}
Salvatore~R. Manmana, Alejandro Muramatsu, and Reinhard~M. Noack.
\newblock Time evolution of one-dimensional quantum many body systems.
\newblock {\em AIP Conf. Proc.}, 789:269, 2005.

\bibitem{Vidal2004}
G.~Vidal.
\newblock Efficient simulation of one-dimensional quantum many-body systems.
\newblock {\em Phys.~ Rev.~ Lett.}, 93:040502, 2004.

\bibitem{WhiteFeiguin2004}
S.~R. White and A.~E. Feiguin.
\newblock Real time evolution using the density matrix renormalization group.
\newblock {\em Phys.~ Rev.~ Lett.}, 93:076401, 2004.

\bibitem{DaleyVidal2004}
A.~J. Daley, C.~Kollath, U.~Schollw\"ock, and G.~Vidal.
\newblock Time-dependent density-matrix renormalization-group using adaptive
  effective {H}ilbert spaces.
\newblock {\em J.~ Stat.~ Mech.: Theor.~ Exp.~}, P04005, 2004.

\bibitem{AffleckLudwig1991}
Ian Affleck and Andreas W.~W. Ludwig.
\newblock Universal noninteger \char96{}\char96{}ground-state
  degeneracy\char39{}\char39{} in critical quantum systems.
\newblock {\em Phys. Rev. Lett.}, 67(2):161--164, Jul 1991.

\bibitem{LaflorencieAffleck2006}
Nicolas Laflorencie, Erik~S. Sorensen, Ming-Shyang Chang, and Ian Affleck.
\newblock Boundary effects in the critical scaling of entanglement entropy in
  1d systems.
\newblock {\em Physical Review Letters}, 96(10):100603, 2006.

\bibitem{VidalKitaev2003}
G.~Vidal, J.~I. Latorre, E.~Rico, and A.~Kitaev.
\newblock Entanglement in quantum critical phenomena.
\newblock {\em Phys.~ Rev.~ Lett.}, 90:227902, 2003.

\bibitem{KuehnerMonien2000}
T.~D. K\"uhner, S.~R. White, and H.~Monien.
\newblock One-dimensional {B}ose-{H}ubbard model with nearest-neighbor
  interaction.
\newblock {\em Phys.~Rev.~B}, 61(18):12474, 2000.

\bibitem{ZakrzewskiDelande2008}
Jakub Zakrzewski and Dominique Delande.
\newblock Accurate determination of the superfluid-insulator transition in the
  one-dimensional bose-hubbard model.
\newblock {\em arXiv:0701739}, 2007.

\bibitem{BuonsanteVezzani2007}
P.~Buonsante and A.~Vezzani.
\newblock Ground-state fidelity and bipartite entanglement in the bose-hubbard
  model.
\newblock {\em Physical Review Letters}, 98(11):110601, 2007.

\bibitem{Lieb1963}
E.~H. Lieb.
\newblock Exact analysis of an interacting {B}ose gas.~ {II}.~ the excitation
  spectrum.
\newblock {\em Phys.~ Rev.}, 130:1616, 1963.

\bibitem{KollathZwerger2004}
C.~Kollath, U.~Schollw\"ock, J.~von Delft, and W.~Zwerger.
\newblock One-dimensional density waves of ultracold bosons in an optical
  lattice.
\newblock {\em Phys.~ Rev.~ A}, 71:053606, 2005.

\bibitem{AltmanAuerbach2002}
E.~Altman and A.~Auerbach.
\newblock Oscillating superfluidity of bosons in optical lattices.
\newblock {\em Phys.~ Rev.~ Lett.}, 89:250404, 2002.

\bibitem{HuberBlatter2007}
S.~D. Huber, E.~Altman, H.~P. Buchler, and G.~Blatter.
\newblock Dynamical properties of ultracold bosons in an optical lattice.
\newblock {\em Physical Review B (Condensed Matter and Materials Physics)},
  75(8):085106, 2007.

\bibitem{Cazalilla2004}
M.~A. Cazalilla.
\newblock Are the {Tonks} regimes in the continuum and on the lattice truly
  equivalent?
\newblock {\em Phys.~ Rev.~ A}, 70:041604(R), 2004.

\bibitem{GroismanWinter2005}
Berry Groisman, Sandu Popescu, and Andreas Winter.
\newblock Quantum, classical, and total amount of correlations in a quantum
  state.
\newblock {\em Physical Review A (Atomic, Molecular, and Optical Physics)},
  72(3):032317, 2005.

\bibitem{AmicoMassimapalma2004}
Luigi Amico, Andreas Osterloh, Francesco Plastina, Rosario Fazio, and
  G.~Massimo Palma.
\newblock Dynamics of entanglement in one-dimensional spin systems.
\newblock {\em Physical Review A (Atomic, Molecular, and Optical Physics)},
  69(2):022304, 2004.

\bibitem{ChiaraFazio2006}
G.~De Chiara, S.~Montangero, P.~Calabrese, and R.~Fazio.
\newblock Entanglement entropy dynamics in {H}eisenberg chains.
\newblock {\em J. Stat. Mech.}, page P03001, 2006.

\bibitem{BarmettlerGritsev2008}
Peter Barmettler, Ana~Maria Rey, Eugene Demler, Mikhail~D. Lukin, Immanuel
  Bloch, and Vladimir Gritsev.
\newblock Quantum many-body dynamics of coupled double-well superlattices.
\newblock {\em arXiv:0803.1643}, 2008.

\bibitem{WolfCirac2008}
Michael~M. Wolf, Frank Verstraete, Matthew~B. Hastings, and J.~Ignacio Cirac.
\newblock Area laws in quantum systems: mutual information and correlations.
\newblock {\em Phys. Rev. Lett.}, 100:070502, 2008.

\end{thebibliography}


\end{document}